\documentclass[10pt]{article}
\usepackage{graphicx}

\usepackage{epsfig}
\usepackage{latexsym}
\usepackage{amsmath}

\begin{document}

\title{\Large \bf Breakdown of Benford's Law for  Birth Data
 }

\author{ \large \bf M. Ausloos$^{1,2,}$\footnote {marcel.ausloos@ulg.ac.be,} ,    C. Herteliu$^{3, }$\footnote {claudiu.herteliu@gmail.com} , B. Ileanu$^{3, }$\footnote {ileanub@yahoo.com} ,  
 \\ \\  $^1$ e-Humanities Group, KNAW,\\Joan Muyskenweg 25, 1096 CJ Amsterdam, The Netherlands \\
\\ $^2$ GRAPES, rue de la Belle Jardiniere, B-4031 Liege, \\Federation Wallonie-Bruxelles, Belgium
\\ 
\\$^3$ The Bucharest University of Economic Studies,  Bucharest, Romania \\ }

\maketitle

  \begin{abstract}
Long birth time series for  Romania  are investigated from Benford's law point of view, distinguishing  between families with a   religious  (Orthodox and Non-Orthodox)  affiliation. The data  extend   from  Jan. 01, 1905  till   Dec. 31, 2001, i.e. over 97 years or 35 429 days. The results  point to a drastic breakdown of Benford's law.
Some interpretation is proposed, based on the statistical aspects due to  
 population sizes, rather than on human thought constraints when the law breakdown is usually expected.
Benford's law breakdown  clearly points to  natural causes.
\end{abstract}


 \maketitle

\section{  	Introduction }\label{introduction}

Newcomb   \cite{ref[1]}  and later   Benford  \cite{ref[2]} observed  that the occurrence of significant digits in many data sets is $not$ uniform but tends to follow a logarithmic distribution such that the smaller digits appear
as  the first significant digits more frequently than the larger ones, i.e.,  
\begin{equation}\label{Beneq1}
N_{d}= N\; log_{10}(1+\frac{1}{d}), \;\;\; d = 1, 2, 3, . . . , 9
\end{equation}
where $N$ is the total number of considered 1-st digits for  checking the law, in short, the number of data points,  and  $N_{d}$ is the number of the observed integer $d$ ( $= 1, 2, 3, . . . , 9$). 
Usually, it seems that Benford's law breaks down  when there is human manipulation  or control (in various ways) of the data.   

The literature on the subject is enormous \cite{0607168befordlawbiblio,beebe2013bibliography} and not all papers can be quoted here. A few of socio-econo-statistical physics papers of  interests are  pointed out in Section  \ref{stateoftheart}.

  In this paper,  our goal is  to  investigate  whether  Benford's law holds, on long birth time series,  distinguishing  between the religious adhesion  (Eastern Orthodox or not)  of families in  Romania for  a time interval extending from   Jan. 01, 1905  till   Dec. 31, 2001, i.e. over 97 years or 35 429 days. The results  point to a drastic breakdown of Benford's law.
Some interpretation is proposed, based on the statistical aspects due to  
 population sizes, rather than on human thought constraints.

In Section  \ref{sec:dataset},  the data acquisition  is recalled.  It leads to a set of  time series. 
 The data of interest  are displayed   through histograms and  discussed   following   a statistical analysis in
 Section  \ref{histodailyrange}.
 All Benford law tests are found in Section  \ref{sec:BL}, in particular with 
 a test of Benford's law for the  1st and 2nd digits of the time series of the    daily birth number of babies  in Romania,   distinguishing between Orthodox  and Non-Orthodox families. 
 
 Since the results  point to a drastic breakdown of Benford's law, a
   discussion of the findings, followed by an explanation,  is found in Section  \ref{discussion}.  

Section  \ref {sec:conclusions} serves for a conclusion emphasizing (i) the interest of such  a data  study along Benford's law concepts, and (ii) the complexity of studying a community, and its religiosity,  through its  baby birth  history.  

\section{Benford's law: a short state of the art literature review}\label{stateoftheart}

 The applications of Benford's law are too numerous to be all quoted here  \cite{0607168befordlawbiblio,beebe2013bibliography}. Nevertheless, for shining some light on the subject, we point to those Benford's law showing  detection of data anomalies in  actuarial and financial  cases \cite{Nigrini96}   
  -\cite{GER12.11.243}, and  also in political cases \cite{ref[21],Roukema} and surveys \cite{JudgeSchechter}. 

Beside these  fields of applications,  Benford's law  has been applied in less dwelled subjects,  e.g., when discussing the appearance of numbers on the internet \cite{[120]}, or recently,
 \cite{SCIM98.14.173Alves}  for  comparing articles of scientific journals. The law has been suggested to be also useful for   optimizing the size of computer files \cite{[18]} or for enhancing computing speed \cite{[19]}.
 
In biological sciences,  Benford's law has been utilized  to check the veracity of the data on clinical trials    \cite{[15]} and discovery of drugs   \cite{[16]},  and in the study of diseases and genes \cite{[17]}.  Similarly, 
in physics,  Benford's law has been used to detect data anomalies in   numerical data on physical constants \cite{[10]}, atomic spectra  \cite{[11],EJP14}, decay width of hadrons \cite{[12]}, magnitude and depth of earthquakes \cite{[13]}, while   in astrophysics \cite{IJMPC17.06benfordastro},  for the mantissa distribution of pulsars \cite{[14]} or the distances of galaxies and stars \cite{Leontsinis}.
 
In econophysics,  the yearly financial reports  of  the Belgian Antoinist community, income and expenses,  were  examined along the so-called Benford's law in order to detect any wrongdoing in the finance of such a sometimes called religious sect.  Nothing anomalous was found. Note that    an imperfect  ("generalized") Benford's law-like form,  better suited for   distributions presenting a minimum at some intermediate digit,  was presented in \cite{clippe2012benford}.

In socio-physics, closely related to our subject, some data  analysis distinguishing between religious adhesion,   Mir \cite{TAMir012,TAMir014} investigated  whether regularities or anomalies exist in numerical data on the country-wise adherent distribution of seven major world religions  along Benford's law  .

For completeness, basic  (mathematical) considerations are found in
    \cite{Raimi76}-\cite{PS8.11.1}.

\section{Data}\label{sec:dataset}  

  \begin{figure}
\centering
  \includegraphics [height=9.5cm,width=12.5cm]  {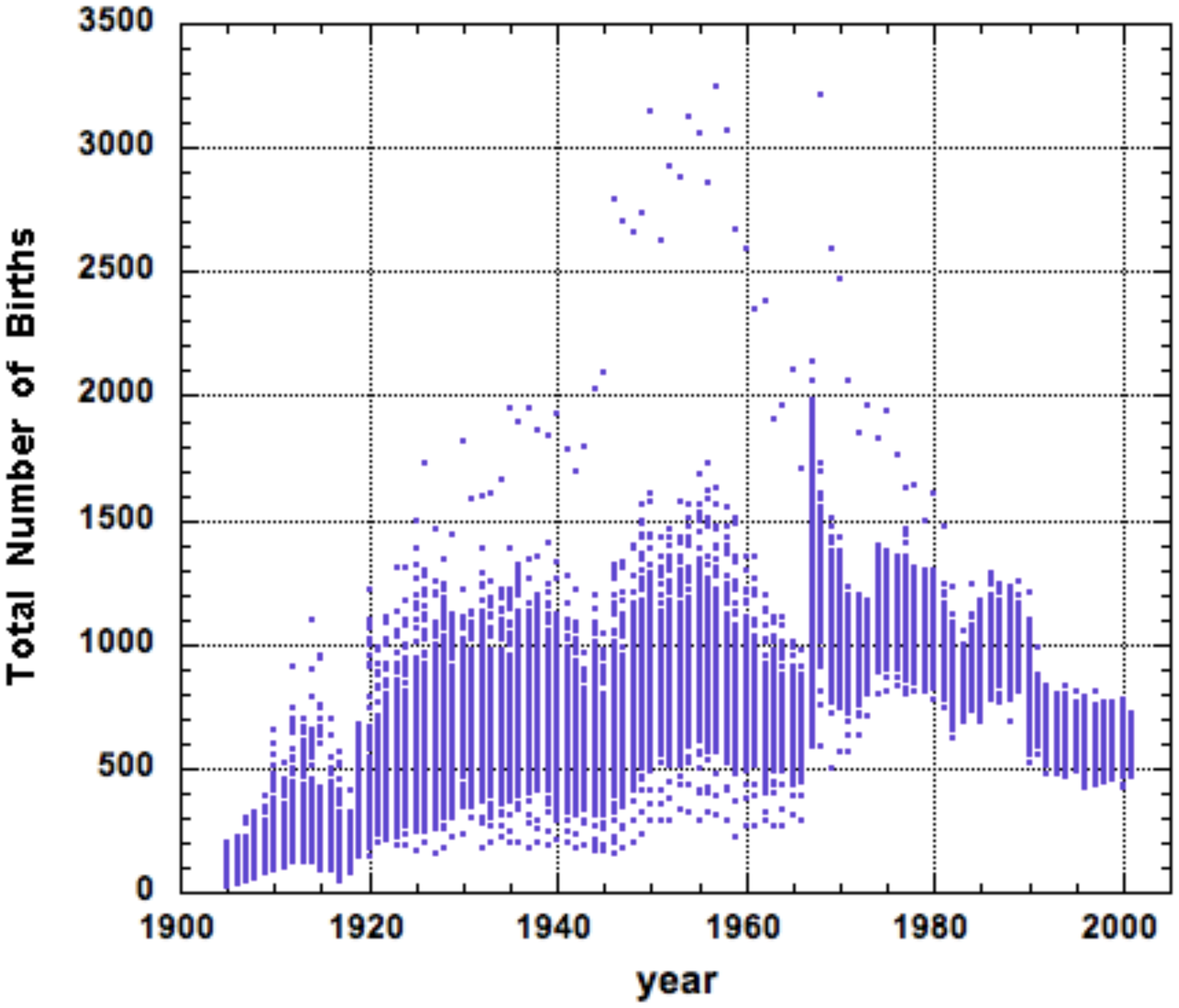}  
 \caption   {   
Number of babies born  in  Romania,  between Jan. 01, 1905 and Dec. 31, 2001, during each indicated year, and still alive on that day    according to the 2002 census; each   point corresponds to a specific number occurrence.  Several data points overlap because  the number of born babies occurs to be the same on various days during the year. }   \label{Plot27TotalNbirthsperyr} 
\end{figure}

The data  were obtained from  1992 and 2002 censuses by the Romanian National Institute of Statistics (NIS).  
 The data of interest  pertain  to the record of  the total number of births in Romania for 
persons still alive at the  1992  and 2002 census reference points. In this respect, the data might rather be called "survival  occurrence from birth date".   
The fact that the true daily birth data are not known to us is irrelevant for the present considerations. Thereafter, we will  use,  for conciseness,  the vocabulary  "born  per day" instead of the   "number of births on a given day   for persons still alive in 1991  and 2001, respectively, for the 1992  and 2002 census". 

The data census allows to distinguish the population under  various criteria, e.g. the religious adhesion. Such data must be   taken with caution \cite{TAMir012,TAMir014,religion1,religion2}, but the orders of magnitude are  usually trustworthy \cite{hayw99}-\cite{religionPRE}.
The  most important community,  from  the religion point of view,  according to the census sources  is the Eastern-Orthodox (86.8\%). The  so called here below  "Non-Orthodox families"  (13.2\%) are mainly made of Roman-Catholics (4.7\%), Reformed Church (3.2\%), Pentecostals (1.5\%), Greek-Catholics (0.9\%), Baptists (0.6\%), Seventh Day Adventists (0.4\%),  Moslems (0.3\%), Unitarians (0.3 \%), Lutherans (0.3\%), Evangelicals (0.2\%) and  Old Rite Christians (0.2\%).  Other denominations (0.6\%), including atheists,  have  each a smaller   size.  

  \subsection{Time series }\label{timeseries}
  
   Thus, we have some information on births  in Romania from  Jan. 01, 1905  till   Dec. 31, 2001, i.e. 97 years or 35 429 days,
see Fig. \ref{Plot27TotalNbirthsperyr}.	 Note that several data points overlap in the display, because  the number of born babies occurs to be the same on various days during the year.  It is also observed that  several occurrences  seem to exist as  outliers.  We have kept the official data unchanged.

\begin{figure}
\centering
  \includegraphics [height=8.5cm,width=12.5cm] {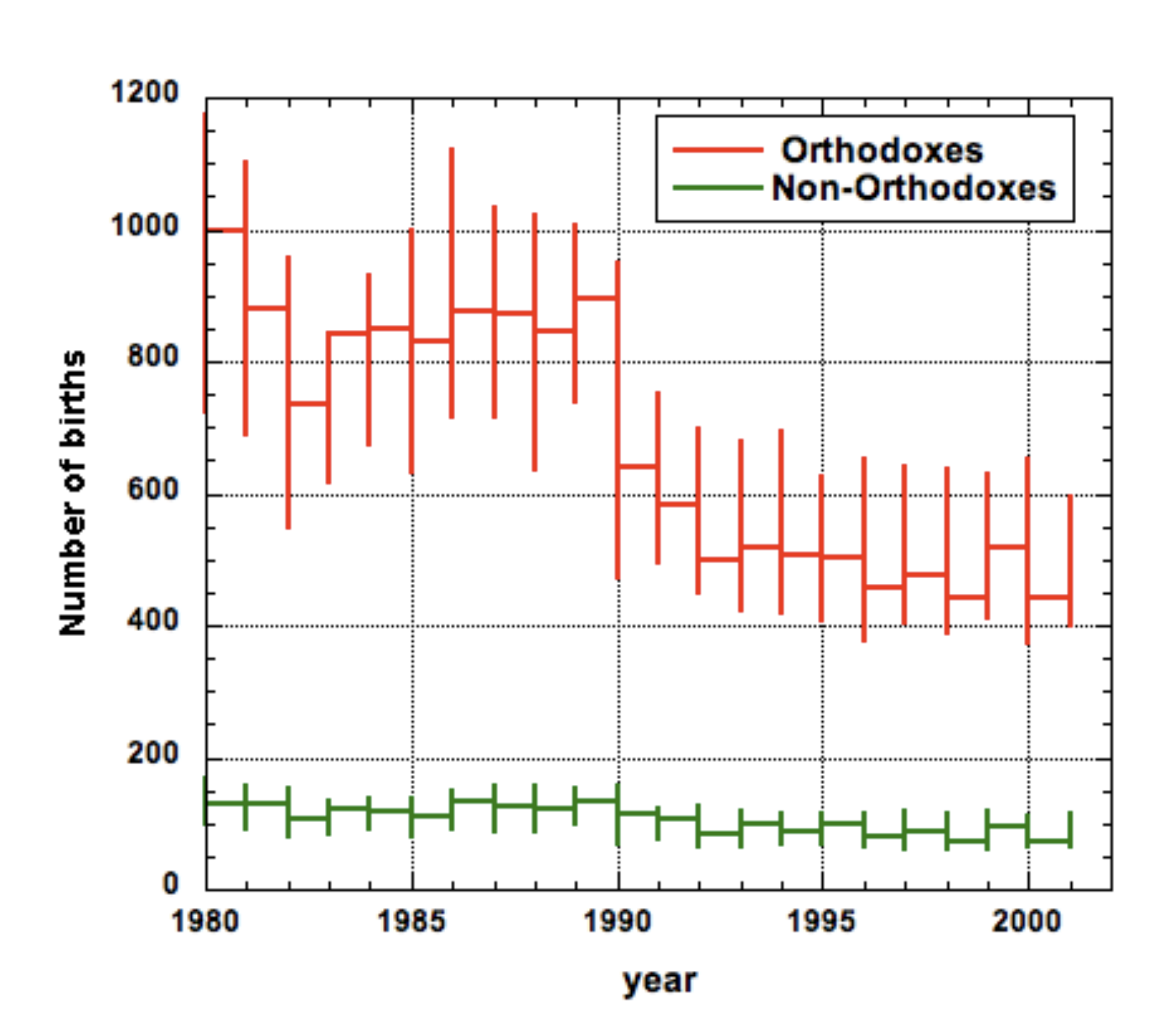} 
 \caption   {   
Number of babies, still  alive in   2001, according to the 2002 census, born in  Orthodox  (red bars) families and  Non-Orthodox families (green bars),  illustrated in the form of  a  yearly step (high-low)  time series  in [1982-2001] giving the yearly range. The horizontal bars indicate the mean of the corresponding number of babies in the first year of  two consecutive years.}   \label{Plot23data28TotOxNOx} 
\end{figure}

Part of the resulting time series for the number of born babies are shown in Fig.  \ref{Plot23data28TotOxNOx} for the case of Non-Orthodox and Orthodox families,  for the last 20 years, [1982-2001], of the examined whole time interval.  
These short time interval examples have been selected  in order to show  that nothing very drastic seemed to have occurred then at a time of turmoil. Each horizontal bar between two consecutive years refers to the mean of the  former year. Nevertheless,   a large daily variation, the height of the bar,  is seen within each year. However,   the orders of magnitude are barely changed. Nevertheless, observe the respective orders of magnitude on the $ y$-axes for  both  types of families: the ratio between Non-Orthodox and Orthodox birth number in such families is about   15\%. This is in agreement with  the  record for the whole population 1992 census.   The same  ratio was also recorded for the 2002 census.  However, the downward jump occurring in 1990, after the Berlin wall  fall and communist regime fall in Romania, is  well marked in both  cases.  This is more strongly so for the case of Orthodox families,  Fig.  \ref{Plot23data28TotOxNOx}. This   can be interpreted as due to  some   "woman liberalization condition", e.g. to a spontaneous relaxation with respect to Ceausescu Oct. 01, 1966 decree ($\#$770) forbidding abortion. In this respect, note the remarkable peak in 1967, on  Fig. \ref{Plot27TotalNbirthsperyr}:    
the cohort born in 1967 doubled compared to that in the previous year.

  \begin{table}
   \begin{center}
\begin{tabular}[t]{ccccc }
  \hline
   $ $   &Total &Orthodoxes &Non-Orthodoxes&  \\
\hline
Minimum		&19 	&	14 	&	2 	\\			
Maximum	&	3249 	&	2942 	&	377 	\\			
Total (x 10$^6$)	&	24.947 &	21.630 	&	3.317  \\			
Mean ($\mu$)	&	704.14	&	610.51	&	93.631	\\			
Median ($m$)	&	714 &	617 	&	97 	\\
Mode($M$)	&	628.3&	626.2 	&	91.05 	\\			
RMS	&	769.31	&	670.59	&	99.982	\\			
Std. Dev.  ($\sigma$)	&	309.89	&	277.43	&	35.065	\\			
Var. &	96030	&	76967	&	1229.5	\\			
Std. Err. 	&	1.6464	&	1.4739	&	0.18629	\\			
Skewness &	0.049587	&	0.096844	&	-0.17298	\\			
Kurtosis 	&	1.1619	&	1.1326	&	1.5394	\\	

($\mu/\sigma$)	 & 2.272 & 2.201  &2.670 \\\hline 
\end{tabular}
   \caption{Rounded statistical characteristics of the daily  number of persons born  between Jan. 01, 1905 and   Dec. 31, 2001, 
having survived till the 2002 census,  for the whole Romania, and also distinguishing between babies born in  either Orthodox  or  Non-Orthodox religiously oriented families.    }\label{Tablestat}
\end{center}
 \end{table}

  \subsection{Histograms daily range}\label{histodailyrange}

The number of babies born   and still alive   depends on the year and the day. The discussion of  such a time dependence is outside the scope of the  present study.  

\begin{figure}
\centering
  \includegraphics [height=7.5cm,width=12.5cm]{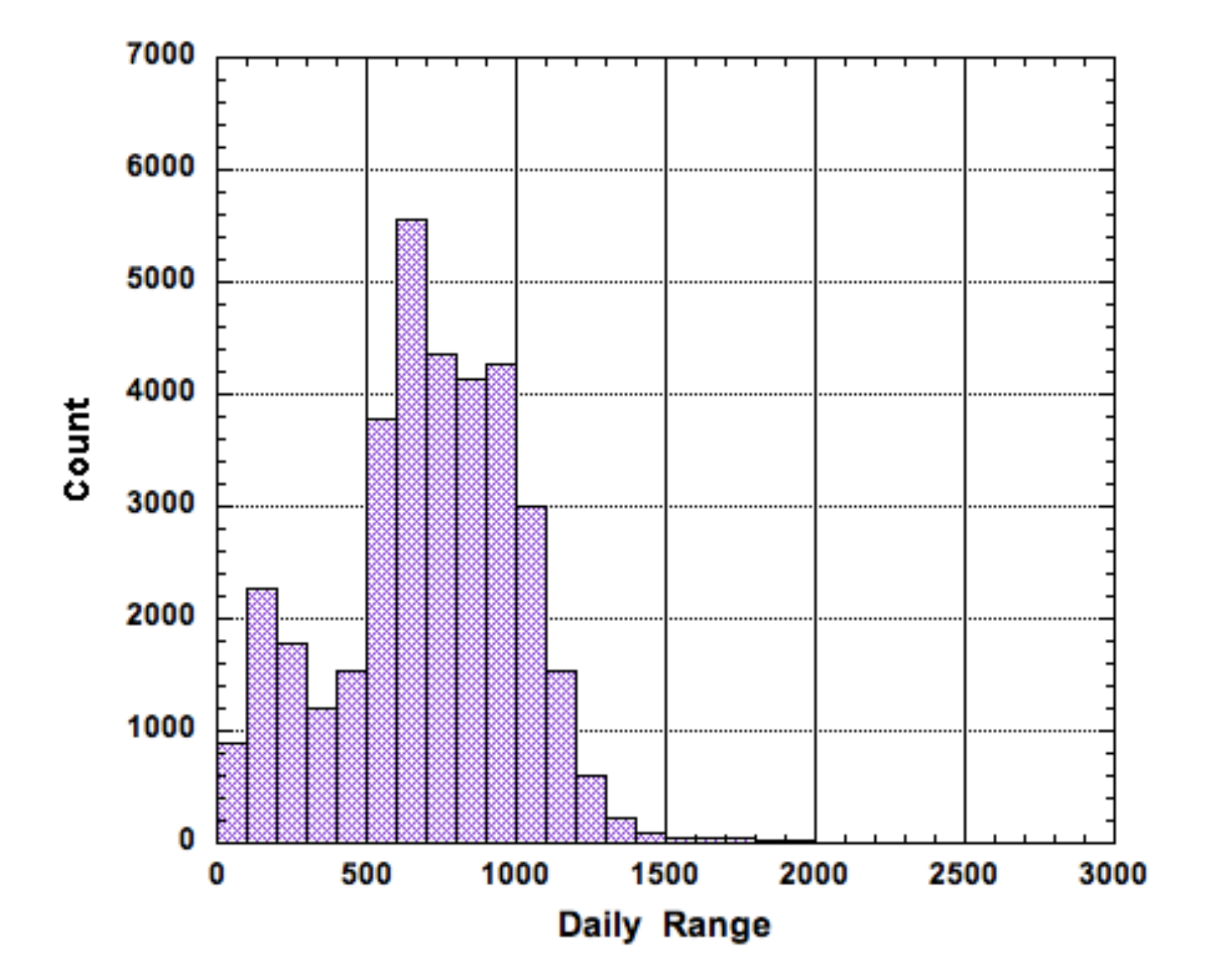} 
 \caption   {   
Histogram of the number of babies born  per day in whole romanian families during the time interval examined in the text}   \label{Screenshotplot0} 
\end{figure}

\begin{figure}
\centering
  \includegraphics [height=7.5cm,width=12.5cm]{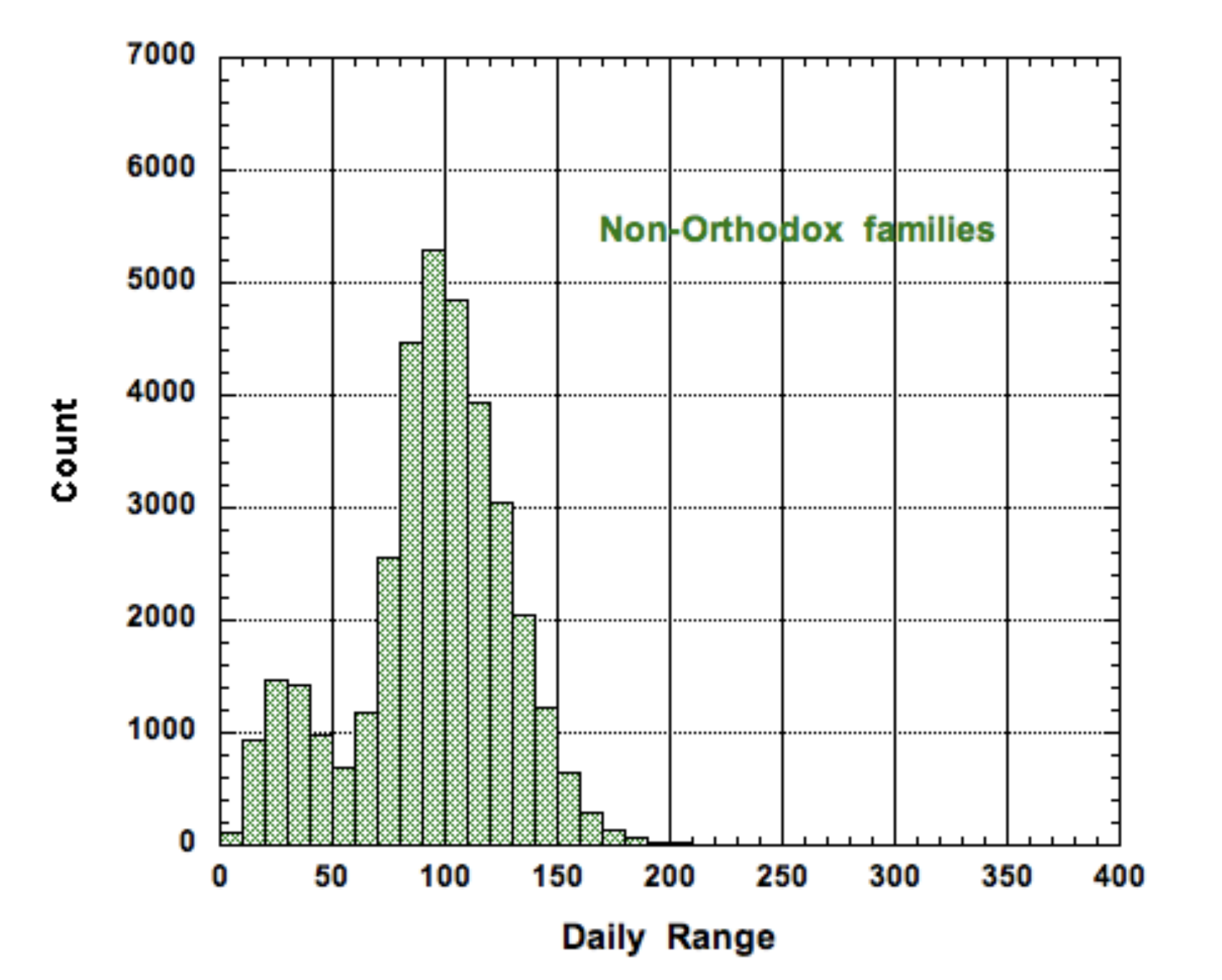}
 \caption   {  
Histogram of the number of babies born  per day in Non-Orthodox families}   \label{Plot119histoBNOx} 
\end{figure}

\begin{figure}
\centering
  \includegraphics [height=7.5cm,width=12.5cm] {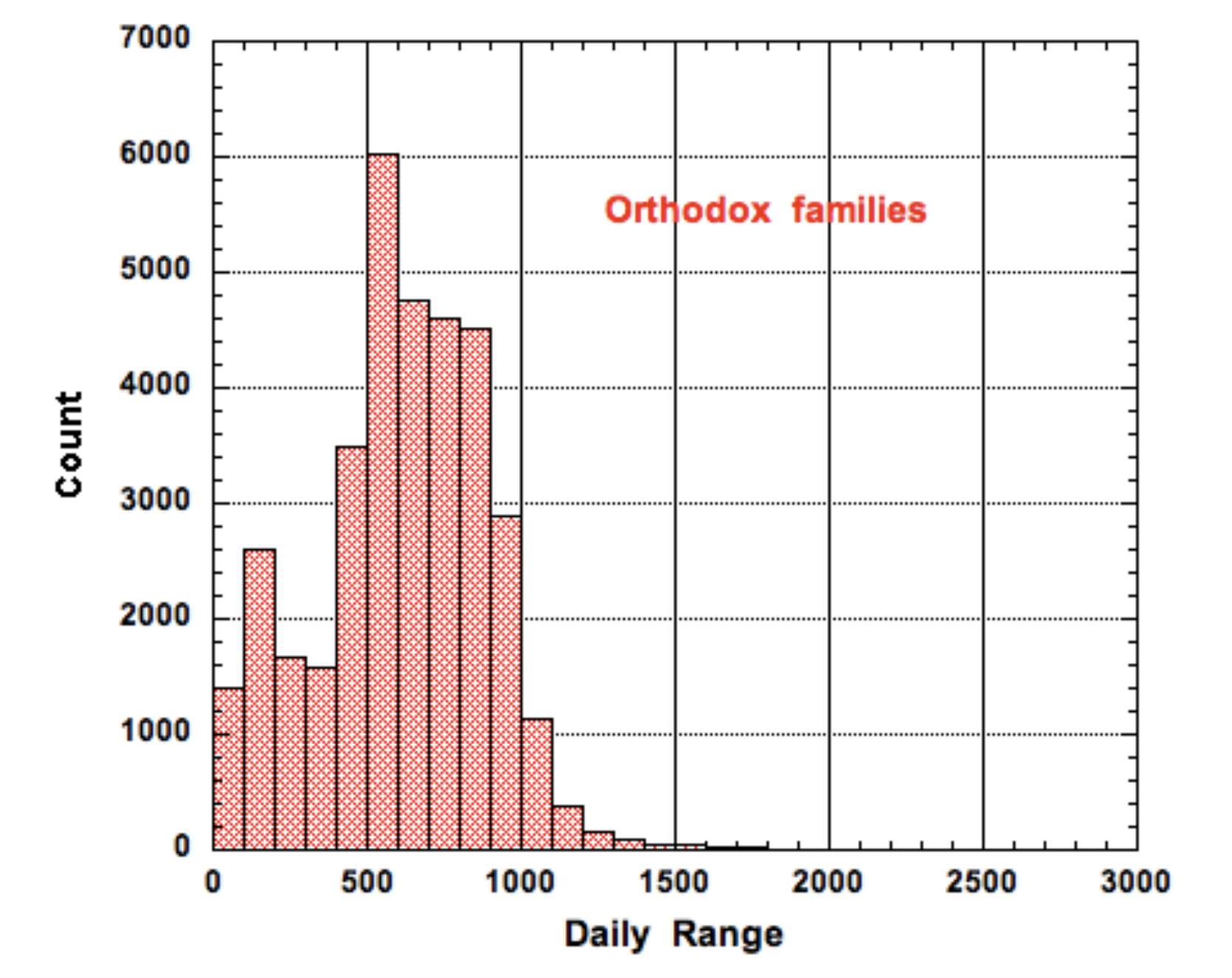} 
 \caption   {   
Histogram of the number of babies born  per day in  Orthodox families }   \label{Plot118histoBOx} 
\end{figure}

 Fig.  \ref{Screenshotplot0}  is a histogram of the number of babies born per day in Orthodox and Non-Orthodox families in the examined time interval.  The   distribution for  Orthodox  and Non-Orthodox families, respectively, is given in Fig.  \ref{Plot119histoBNOx} and  Fig.  \ref{Plot118histoBOx} also under a histogram form. 
 
  More statistical considerations are given in Table \ref{Tablestat}. The minima and the maxima i.e. the  daily ranges   are [14; 2942] and [2; 377],  respectively,  for  Orthodox  and    Non-Orthodox families.   The range width ratio  itself  is  $\simeq$ 7, rather larger than the population census ratio.

Note that for the considered century time interval,  the  ratio in the number of babies born in  Orthodox  and  in Non-Orthodox families is  about  6.5 ($\simeq 21630/3317$). This corresponds well to the population ratio  census, reporting  a  6.6  ratio  between such families.   This denotes a lower fertility for Orthodox families \cite{isaicManiuHerteliu}. 
 
An interesting value is the mean to standard deviation ratio which allows to admit that the mean value is that of a single peak or so distribution. It indicates a more peaky distribution for the Non-Orthodox case.   In fact, for this community, the skewness is negative.  The  kurtosis is  positive in all cases. 
 
\begin{figure}
\centering
  \includegraphics [height=7.5cm,width=12.5cm]{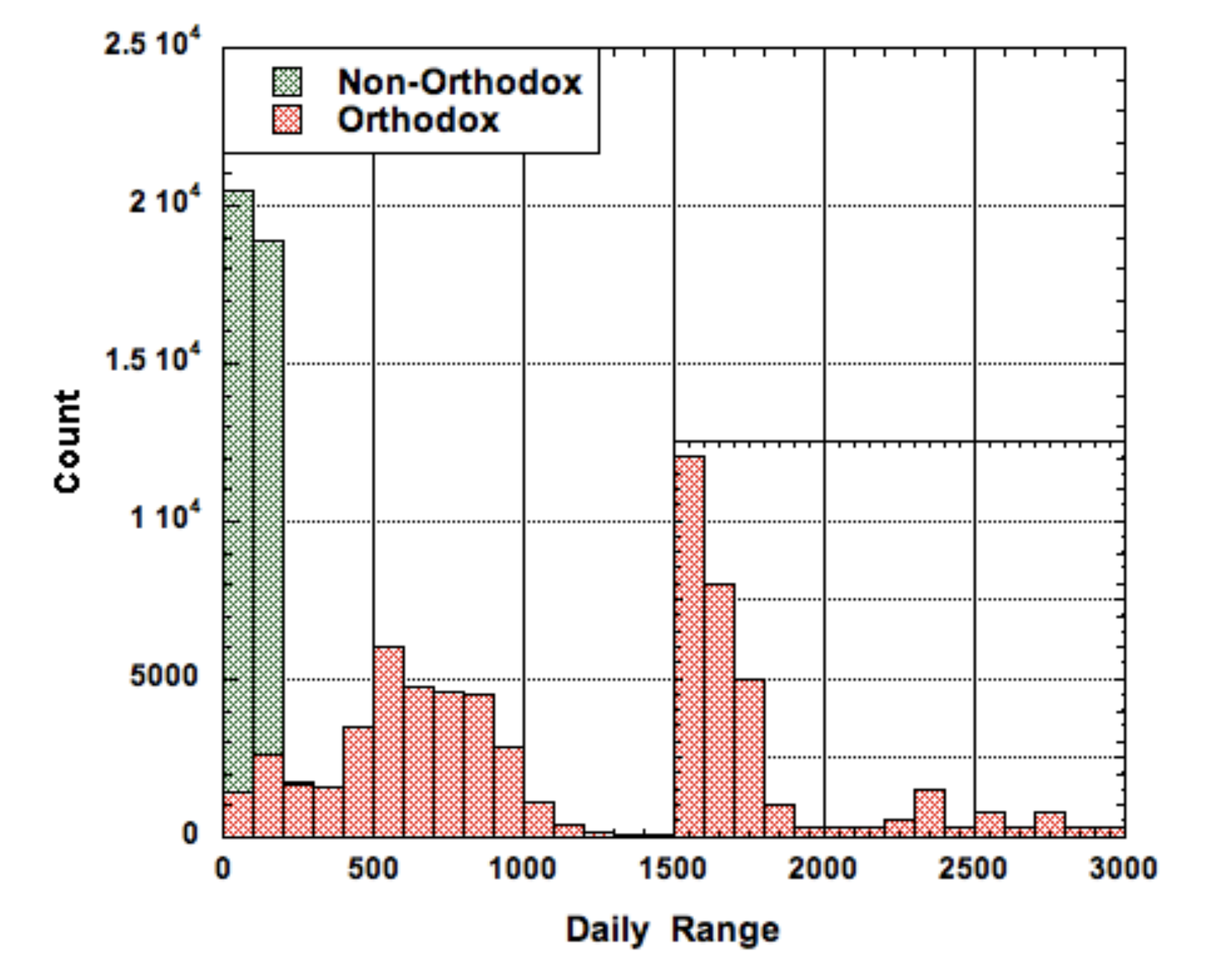}
 \caption   {   
 Histogram of the number of babies born  per day in Orthodox  and Non-Orthodox families. The regime of high daily range (above 1500 daily range) inset has a  250  times  scale increased $y$-axis 
}   \label{Plot120histoBOxBNOxmx} 
\end{figure}

Therefore a specific comment is in order: from Fig.  \ref{Plot120histoBOxBNOxmx}, i.e.  the  
 histogram of the number of babies born  per day in Orthodox  and Non-Orthodox families, one reads  that there are usually  a few  (less than 100) Non-Orthodox babies born per day. Therefore, it can be concluded that  they are rather evenly distributed during the year. However, the Orthodox families   have   babies in short time intervals during the year. Our interpretation   points to religious constraints.  Indeed, the Orthodox church rules request to have less sexual activity  in periods before Easter and Christmas. Due to this    restriction on sexual activity, this is implying that the  average number of babies per day  appears  necessarily  larger in the Orthodox case.

\section{Benford's laws tests }\label{sec:BL}

Due to the range in the number of births, it seems interesting to  test Benford's law, not only for the first reported digits, but also for the second digit. In practice, applications of Benford's law for fraud detection routinely use more than the first digit \cite{Nigrini96}.

\subsection{Benford's law 2nd digit }\label{sec:BL2}
Indeed, the above  Eq.(1)  can be extended to forecast how many times any digit, or  also any combination, should be found at some rank in the  string of digits \cite{Nigrini96},  requoted by \cite{Durtschietal04}.  In the latter case, the  "0" has to be taken into account.

The  probability of encountering a number starting  a string of digits with the digit $n$ is given by 
\begin{equation}\label{eq2}
  \pi\;=
  \log_{10}\left( \frac{n+1}{n}\right). \end{equation}
Thus, the probability that $d$ ($d$ = 0, 1, ..., 9) is encountered as the $n$-th ($n$ $ >$  1) digit is
\begin{equation}\label{eq3}
 p_d(n)\;=
 log_{10}\left[\;\prod_{k=10^{n-2}}^{10^{n-1}-1} (\frac{10k+d+1}{10k+d})\right].
\end{equation}

 For instance, the probability that a  "0" is encountered,  \cite{Nigrini96},  as the second digit is 
  \begin{equation}\label{eq4}
p_0(2)\;=
 \log_{10}\left(1 +\frac{1}{90}\right) \approx  0.1197. \end{equation}
 The variation is very smooth.  It is easy to find that the probability that a  "1"  or a  "9" is encountered as the second digit is  $\approx$  0.1139 and 0.0850, respectively.

 \begin{figure}
\centering
  \includegraphics [height=6.5cm,width=9.5cm]{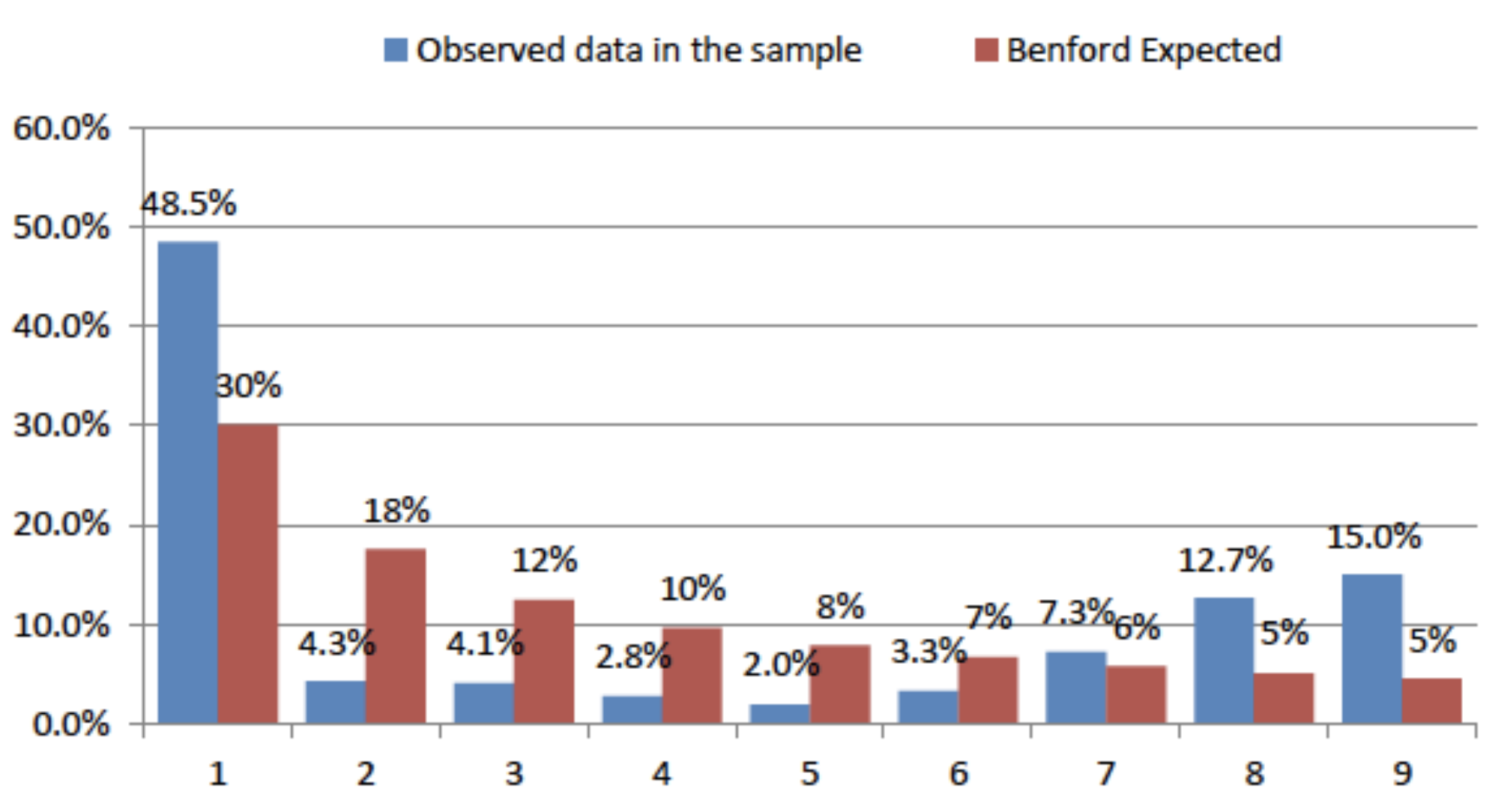}
 \caption   {  
Comparing   Non-Orthodox births/day first digit frequency and expected frequencies according to Benford's law.}
\label{F1Screen shot}
\end{figure}

  \begin{figure}
\centering
  \includegraphics [height=6.5cm,width=9.5cm]{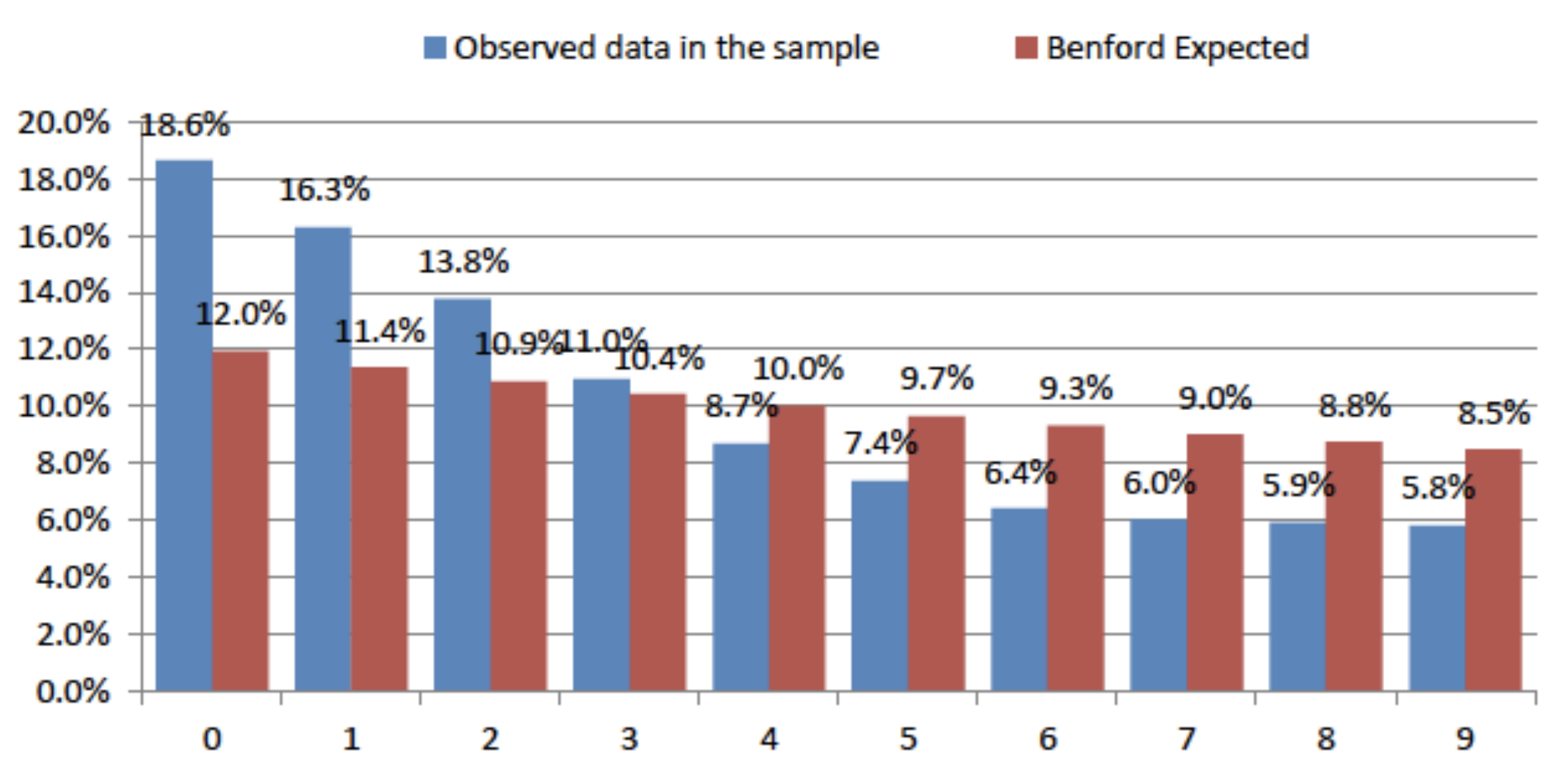}
 \caption   {  
Comparing  Non-Orthodox  births/day second digit frequency and expected frequencies according to Benford's law.}
\label{F3Screen shot}
\end{figure}

  \begin{figure}
\centering
  \includegraphics [height=6.5cm,width=9.5cm]{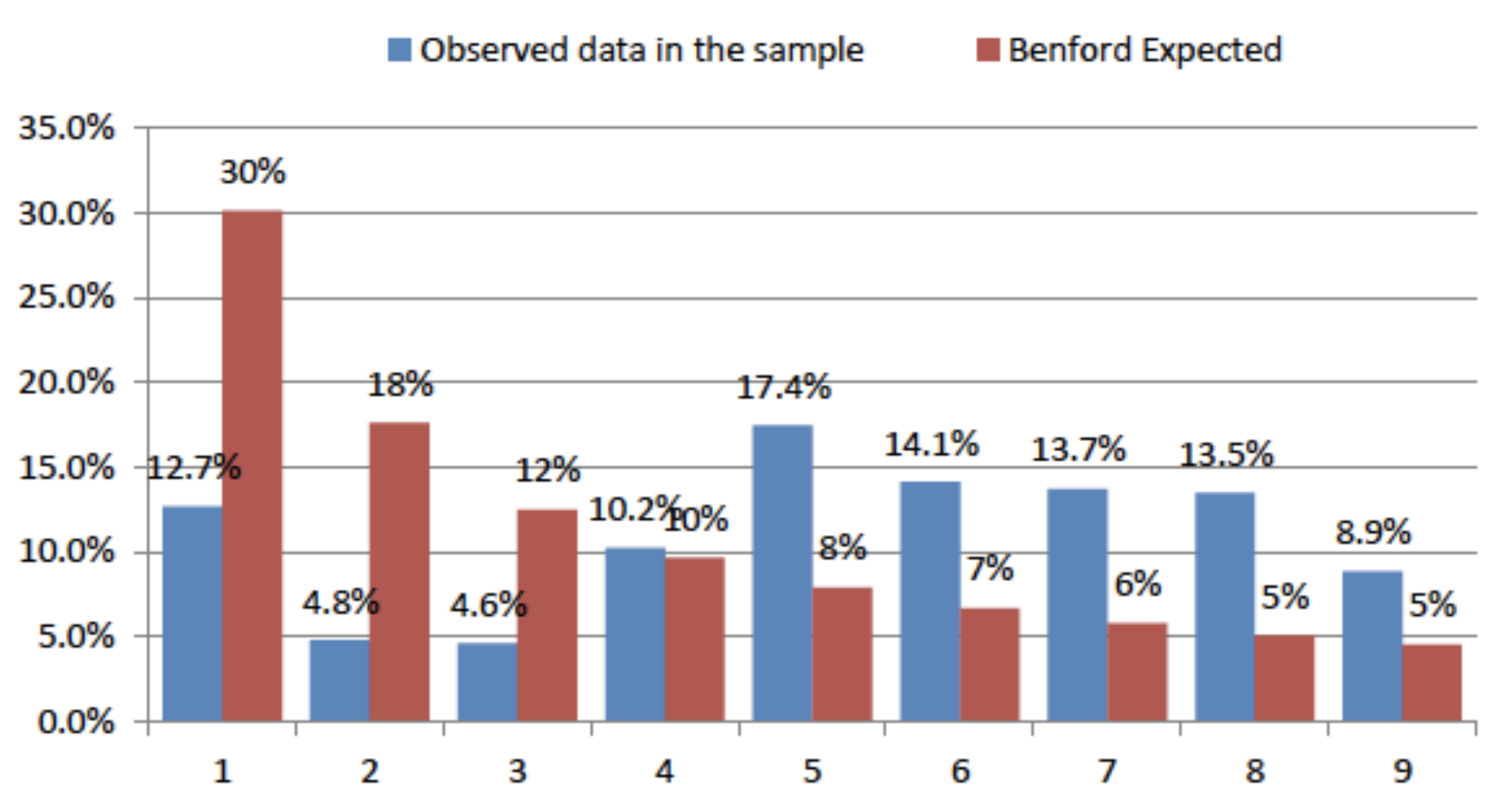}
 \caption   {  
Comparing   Orthodox births/day  first digit  frequency and expected frequencies according to Benford"s law.}
\label{F2Screen shot}
\end{figure}

  \begin{figure}
\centering
  \includegraphics [height=6.5cm,width=9.5cm]{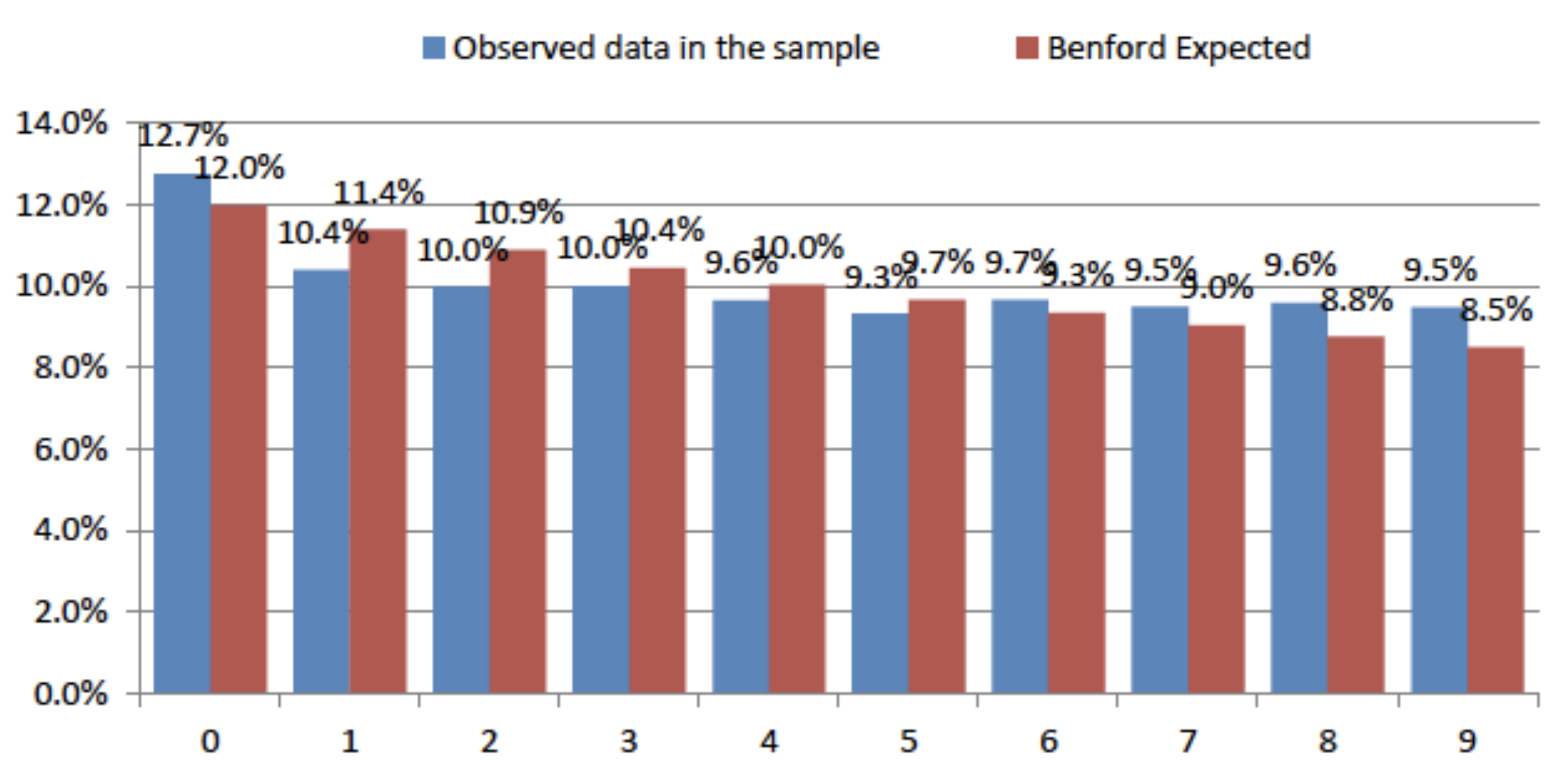}
 \caption   {  
Comparing  Orthodox  births/day  second digit  frequency and expected frequencies according to Benford's law.}
\label{F4Screen shot}
\end{figure}

Fig.   \ref{F1Screen shot} to \ref{F4Screen shot} allow us to compare the  first  and second digit frequencies and  the expected
ones according to Benford's law for the two cases of interest.

 First, it  is  examined whether the distributions of the first digits  match the distribution specified by Benford's law  Eq.(\ref{Beneq1}). Second, it is examined whether the first digits occur as  often as expected  from Eq.(\ref{eq3}) at the second rank. In order to do so a $\chi^2$ test has been used:
 \begin{equation}
 \label{eq:1BLchi2}
\chi^2_{i_1} =\sum_{i=1}^9 \frac{(N_{o,i_1}-N_{e,i_1})^2}{N_{o,i_1}}
\end{equation}
for the first digit and  
 \begin{equation}
 \label{eq:2BLchi2}
\chi^2_{i_2}=\sum_{i=0}^9 \frac{(N_{o,i_2}-N_{e,i_2})^2}{N_{o,i_2}}
\end{equation}
for the second digit, - where $N_{e,i_1}$ and $N_{e,i_2}$ are the theoretically expected values.

The $\chi^2$-distribution ($\chi^2_{(8)}$) with 8 degrees of freedom has a critical value of 15.507 for the 0.05-level  confidence test \cite{Bradley}  while the $\chi^2$--distribution with 9 degrees of freedom ($\chi^2_{9}$) 
has a critical value of 16.91 for the 0.05-level confidence  test.

 \subsection{Application to cases when distinguishing religious adhesions
}\label{BLNOx}
\begin{itemize}
\item
Let us   consider  whether Benford's law is respected for the Non-Orthodox  families.  In this case,   the number of births  can be only 1 digit long, see Table \ref{Tablestat} last column.   There are  108 cases (days); thus 35 321 data entries (days) are only studied, instead of 35 429,  for the second digit aspect.     This  corresponds to a total of  809 babies. 

It is visually observed from Fig.   \ref{F1Screen shot} that Benford's law is unlikely respected;  for the second digit, see  Fig.   \ref{F2Screen shot}.
It is easily calculated that   $\chi^2 = 26 053.68$ and $=171.20$ for the  first and second digit respectively. Thus, a statistically significant difference  is confirmed between  the observed distribution and the theoretical Benford distribution.

\item 
In the case of    Orthodox  families,
a  statistically significant difference is   visually expected   (Fig  \ref{F3Screen shot}) and numerically observed when applying the $\chi^2$ test to  the daily number of birthsÕ first digits.  One obtains  $\chi^2 = 25 760.83$ for the  first   digit, and $\chi^2 = $3888.9 for the second digit (Fig  \ref{F4Screen shot}).
 Thus, there is a major statistically significant difference  in the case of Orthodox
daily number of births.

 \end{itemize}

\section{Discussion }\label{discussion}

 The above  analysis shows a large difference between observation and expectation for the first digit. For the second digit,   
the survey data  are  in rather   close  agreement with  the theoretical distribution. The former observation needs some interpretation.

It seems that Benford's law breakdown can be more easily understood starting from the Non-Orthodox number of births cases,  when observing the histogram data in Fig.  \ref{Plot119histoBNOx}. The peak  count occurs  in the bins 90 to 110.   This fact suggests that the most important birth numbers are 90, 91,  $\dots$, 98, 99, 100, 101,  102, $\dots$,109, 110. Therefore, it can be understood that the most important first digit is 1; it occurs 11 times.  The next most important 1st digit is  9; it occurs 10 times,  as seen in Fig \ref{F1Screen shot}. Furthermore, the most important second digit is  0; it occurs 11 times also.  The other digits are being almost equally possible at the second rank, as seen in Fig.  \ref{F3Screen shot}.

This being understood, a similar reasoning  can be held for the Orthodox number of births cases.  The main peak, see Fig.  \ref{Plot118histoBOx}, extends  from 500 up to 900. Therefore, it is  understandable that the most often occurring first digits, in this case, are 5, 6, 7, 8; see  Fig.  \ref{F2Screen shot}. Concerning the second digit, Fig.  \ref{F4Screen shot},  due to the large range of the main data peak, the second digits are $quasi$ equally likely as observed.  

In both cases, the baby per day peak occurs near the mean or median (see Table 1).  Thus, Benford's law about the 1st digit can  be expected to apply  for Non-Orthodoxes, for which the mean and median are near 100. However, the 1st digit law cannot be valid for Orthodox families, since the mean or median is  near 600.

Therefore, the  origin of the breakdown seems attributable to the  number of couples   "available for" procreating babies during the year.  It has been observed in Table 1  that there are approximately 100  and 600 babies  produced per day, on average, for Non-Orthodox and Orthodox families respectively. We  consider that there are 360 and 300 possible nights, respectively,  for sexual activity due to religious conditions, thereby leading to the  Table 1 data.  Let us assume that procreation occurs for couples with partners who are between 20 and 40 years old.  Assuming much  constancy in the sexual relations during these 20 years,  the respective number of families (or couples) being concerned is about 0.72 10$^6$ and 3.6 10$^6$  respectively. These values seem indeed to be a good order of magnitude for the sexually active and procreative romanian population, per year,  during the last century, and {\it in fine} explain the breakdown of the law for the 1st digit. 

A logical questioning follows: according to Nigrini and Mittermaier  \cite{ref[45]},  Benford's  law should not  (or does not!) apply when human thought is involved (such as supermarket prices and New York Stock Exchange (NYSE) quotation prices)  or when there are  "constraints" as in telephone numbers, lottery numbers, car license plates, or street addresses\footnote{However, Benford considered that the law works for street addresses, item R in TableIV of \cite{ref[2]}, for pre-selected persons}   \cite{Raimi76}.  From these remarks, we should  wonder whether the conception of babies, and their subsequent birth,  results more  from a  Woman-Man thought   than  from some random (or spontaneous) sexual activity. This   interesting question needs further investigation outside the scope of this paper.

 \section{ Conclusions}  \label{sec:conclusions}

Benford's law  universal validity  has always been questioned. It has been explained and/or justified  along various mathematical hypotheses on number occurrences.  However,  when  it is valid in physics and more generally in science is still an open question. Many cases have been discussed as seen in the short list of references   given in the bibliography. There  are  cases, thus data, in which the validity or breakdown can be  fully  proved. Nevertheless,  the causes or origins are debatable. Benford himself examined  20 cases and pointed to   some validity, but with large deviations. The death rate case\footnote{Alas, in \cite{ref[2]}, there is no information on the data origin} came 10th in the validity order. 

We have examined unusually very long data series about birth frequency with survival occurrence up to some census date.  To the question: "Is Benford's  law valid in  the case of birth  data?", our    answer is obviously  "No", according to the $\chi^2$ tests. We have justified the conclusion, finely discussing the survey data.

In conclusion, let us point to three   considerations on the puzzling question:
\begin{itemize}
\item 
 If the numbers under investigation are not entirely random but somehow socially or naturally related, the distribution of the first digit is not uniform, according to  the empirical findings and the theories \cite{Radev}.
 
\item According to Burns  \cite{paper637}, people (largely) follow Benford's  law.   Quoting:  {\it Understanding whether (or when) people follow
BenfordÕs law is important for both practical and theoretical reasons. Practically, the value of BenfordÕs law as a detector of fraud or error is a product of being able to predict when invalid data will nevertheless fit it. Theoretically, it is valuable because it is a precise distribution that every person has had exposure to over their lives. Thus it could be a useful test case for how sensitive people are to a statistical relationship that they are not consciously aware of.}  
 
\item In  Dobrow's recent book \cite{Dobrowbook2013}, it is claimed that  {\it  There are a huge number of data sets which exhibit BenfordÕs law, including street addresses, populations of cities, stock prices, mathematical constants, birth rates, heights of mountains, and line items on tax returns.}


\end{itemize}
 
Three comments must follow, one for each previous item, respectively. On one hand, we have considered   cases of socially or naturally related data distributions, in which people pertain to a distribution to which they are exposed during the whole life. Such a large set of analyzed data is rather rare. On the other hand,  it seems that one should be more restrictive about Benford's law validity in socially or naturally related data. In fact, Raimi \cite{Raimi76} already had seemed to claim the contrary of Dobrow's   (and Benford) about street addresses. The present paper also questions the blind application of Benford's law to birth rates (or survival) data. Benford claimed to have analyzed, among 20 229 real numbers from 20 sources,   one being  "death rates" (item T, in Table 1 of    \cite{ref[2]}),   but without mentioning what   data were  examined, nor the meaning of rate.    It seems that one claims too much concerning Benford's data analysis and possible applications. 

Thus,  we have shown not only  (i) the interest of such  a data  study along Benford's law concepts, but also  (ii) the complexity of studying a community, and its religiosity,  through its  baby birth  history.   

Whence  we  emphasize  that  people are influenced by statistical relationships  due to  their environment, in particular in  socio structural formation, - thereby lowering the   reasoning and decision making  parameter influence to be introduced in socio-physics models.
Therefore, last but not least, we recommend in further work on Benford's law to examine the data distribution at the start of the investigation,  in order to underline so called physical (or more generally, natural) causes at first.

 \begin{flushleft}
{\large \bf Acknowledgment}
\end{flushleft}
 This paper is part of  MA scientific activities in  COST Action TD1210 "Analyzing the dynamics of information and knowledge landscapes'".
 
 This work by CH was co-financed from the European Social Fund through the Sectoral Operational Programme Human Resources Development 2007-2013, project number POSDRU/1.5/S/59184  Performance and excellence in postdoctoral research in Romanian economics science domain.

\end{document}